\documentclass[fleqn,10pt]{wlscirep}
\usepackage[utf8]{inputenc}
\usepackage[T1]{fontenc}
\usepackage{graphicx} 
\usepackage{braket}
\usepackage[title]{appendix}
\renewcommand{\thesection}{\arabic{section}.}
\renewcommand{\thesubsection}{\thesection\arabic{subsection}}

\newcommand{\appendixnumbering}{%
  \renewcommand{\thesection}{\Alph{section}}%
  \renewcommand{\thesubsection}{\thesection.\arabic{subsection}}%
}

\title{Digital Quantum Simulation of Flat-Band and All-Bands-Flat Dynamics for Tunable Quantum Transport}

\author[1,2,*]{Mrinal Kanti Giri}{}
\author[1,3,4$\dagger$]{Pochung Chen}

\affil[1]{Department of Physics, National Tsing Hua University, Hsinchu 30013, Taiwan}
\affil[2]{School of Physical and Mathematical Sciences, Nanyang Technological University, Singapore 637371, Singapore}
\affil[3]{Frontier Center for Theory, Computation, and Data Science, National Tsing Hua University, Hsinchu 30013, Taiwan}
\affil[4]{National Center for Excellence in Quantum Information Science and Engineering‌, National Tsing Hua University, Hsinchu 30013, Taiwan}

\affil[*]{\href{mailto:mrinalphy333@gmail.com}{mrinalphy333@gmail.com}}
\affil[$\dagger$]{\href{mailto:pcchen@phys.nthu.edu.tw}{pcchen@phys.nthu.edu.tw}}

\begin{abstract}
Flat-band systems offer a uniquely powerful tool for quantum control in dynamics due to their characteristic feature of having a dispersionless energy band. Simulating such highly sensitive systems on current digital quantum computers is a challenging task, due to the intrinsic limitations of the noisy intermediate-scale quantum (NISQ) devices. Here we present high-fidelity digital quantum simulations of flat-band (FB) and all-bands-flat (ABF) lattices, using an advanced tensor network based variational optimization approach to compress the circuit depth. With the compressed quantum circuits, we first explore single-particle dynamics and observe two distinct behaviours: strong localization in ABF lattices and delocalization in FB lattices. By integrating FB and ABF lattices into a one-dimensional hybrid structure, we achieve controllable quantum transport, where the ABF lattice acts as a quantum switch. Extending to two-particle dynamics, we show that transport remains controllable by tuning the hopping amplitude alone, even in the presence of interactions. These results establish flat-band engineered systems as a promising pathway for scalable control of quantum transport in emerging quantum technologies, with potential applications in qubit isolation, particle trapping, and state transfer.

\end{abstract}

\begin{document}

\maketitle

\section{Introduction}

Recently, flat-band systems have been of paramount interest in condensed matter physics, particularly in studying transport phenomena, due to their unique electronic properties. These systems feature a dispersionless band structure, where energy remains constant regardless of momentum, resulting in substantial localization effects. In flat-band systems, localization occurs due to destructive interference, forming compact localized states (CLSs) in which wavefunctions are confined to a finite region of the lattice. CLSs exhibit stronger localization compared to Anderson localization~\cite{andersonlocalization1958}, where the wavefunction decays exponentially in the presence of random disorder. Flat bands are particularly intriguing due to their pronounced sensitivity to perturbations such as disorder, interactions, and particle statistics. Owing to these unique properties, flat-band systems provide a rich platform for exploring a wide range of phenomena, including quantum transport~\cite{Danieli2018,PhysRevB.104.085131,Vicencio2021,PhysRevB.103.075415}, localization-delocalization transitions~\cite{PhysRevLett.85.3906,longhi_2021,loc_dloc_2023,loc_dloc_2023_1}, Quantum Hall effects~\cite{FQHE_2011,Sondhi_2013,QAHE_2023}, superconductivity~\cite{PhysRevLett.88.227005,Peotta2015,PhysRevB.83.220503,Shaginyan_2022,Tian2023,PhysRevB.98.155142}, many-body localization~\cite{mbl_flatband2020,mbl_flatband2020_1,mbl_flatband2022} and many more~\cite{Leykam01012018,Danieli2024}.
Several types of flat-band lattices can be realized depending on the underlying geometry~\cite{PhysRevB.98.245116,Pal_2025,Leykam01012018}.
Recent experimental advancements have enabled their implementation across various platforms, including photonic lattices~\cite{PhysRevLett.121.075502,PhysRevLett.126.110501,PhysRevLett.128.256602,Yang2024}, optical lattices~\cite{takahasi2015,PhysRevLett.129.220403,PhysRevLett.126.103601,Chen2025}, electrical circuits~\cite{PhysRevLett.130.206401,PhysRevB.109.075430,PhysRevB.106.104203}, and superconducting circuits~\cite{fb_loc_expt_2023}. These platforms allow for precise control over lattice geometries and the fine-tuning of system parameters, making it possible to achieve flat-band conditions in controlled environments.

\begin{figure}[t]
    \centering
    \includegraphics[width =1.0\linewidth]{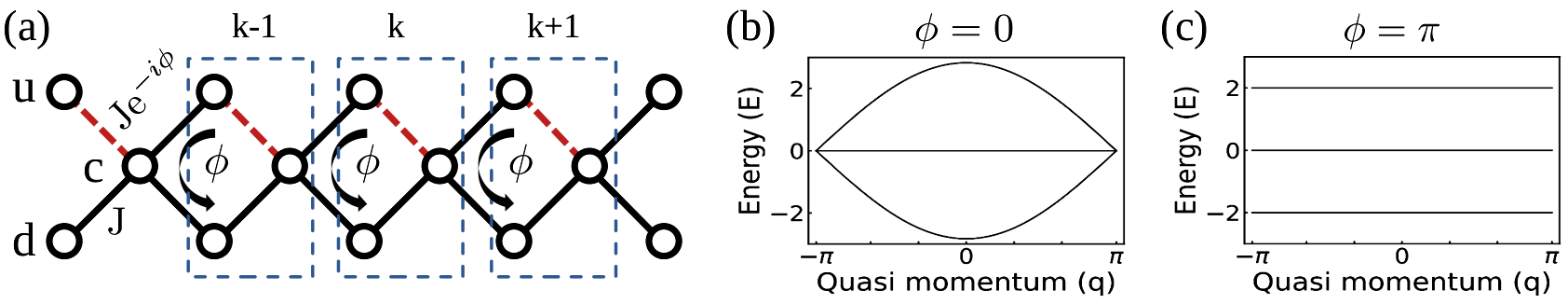}
    \caption{
    (a) Schematic representation of the diamond chain FB lattice, where red dashed line represents the complex hopping. 
    (b) and (c) Energy band diagram for $\phi=0$ and $\phi=\pi$, respectively. The unit cell, outlined in blue dashed lines, consists of the u (up), c (center), and d (down) sites.} 
    \label{fig:flat_lattice}
\end{figure}

In this work, we employ a digital quantum simulator to investigate flat-band dynamics. The goal is twofold. First, we would like to take advantage of the flexibility of digital quantum simulation so that one can explore various different configurations to study corresponding transport properties. Second, we study if quantum circuit compression technique helps to tame the noise and enable reliable digital quantum simulation on current NISQ devices. Specifically, we explore the quantum walk (QW) of single and two particles on the diamond lattice as sketched in Fig.~\ref{fig:flat_lattice}. Here we introduce the magnetic flux $\phi$. By tuning the flux $\phi$, one can reach different configurations which potentially have different transport properties.
When there is no flux insertion, i.e., $\phi=0$, it has one flat band and two dispersive bands - this configuration is referred to as the flat band (FB) lattice. Moreover, two dispersive bands touch the flat band at two ends in the brillouin zone as shown in Fig.~\ref{fig:flat_lattice}(b). Consequently, an initially localized particle will delocalize over time during the quantum walk due to the contribution from the dispersive band. 
At a flux value of $\pi$, all energy bands become flat (see Fig.~\ref{fig:flat_lattice}(c)), forming the all-bands-flat (ABF) lattice. This causes any initially localized particle to exist as a linear combination of CLSs. As a result, the particle's energy remains entirely confined within these states, leading to extreme localization. This phenomenon is known as Aharonov-Bohm (AB) caging, which significantly restricts particle transport and confines the particle to a limited number of lattice sites. The experimental realization of such phenomena has been made possible through artificial gauge fields~\cite{PhysRevLett.121.075502,PhysRevLett.126.103601,PhysRevLett.129.220403,fb_loc_expt_2023}. 
However, experimentally studying the long-time dynamics in highly sensitive flat-band systems remains challenging due to band imperfections, decoherence, experimental instabilities, and measurement limitations.  In contrast, digital quantum computers have emerged as a promising platform for exploring flat-band dynamics. This is because it can provide flexibility in simulating various lattice geometries and precise tuning of lattice parameters. However, current Noisy Intermediate-Scale Quantum (NISQ) computers face significant limitations. Hardware imperfections, decoherence, and limited qubit connectivity present challenges in accurately simulating complex quantum systems over extended periods. To address these limitations, we employ sophisticated tensor network methods to compress the quantum circuit on the classical computer. This enhances computational efficiency and mitigates the noise,  thereby enabling high-fidelity simulations of long-time quantum dynamics. We then utilize the IBM Quantum (IBMQ) platform to simulate the compressed quantum circuits and study quantum walks on the diamond lattice, and benchmark our results.
Following this, we further investigate QWs on the extended ABF lattice, which exhibit strong localization with potential applications such as particle trapping and isolating qubit from crosstalk. We also study QWs in a one-dimensional structure by embedding FB and ABF lattices separately. In this setup, the ABF lattice acts as a quantum switch, enabling dynamic control over particle transport. Finally, we examine the transport of two interacting particles in the presence of an embedded FB or ABF lattice, highlighting the potential of digital quantum computers for simulating and controlling flat-band-engineered systems.

\section{Model and Approach}
We study QW in a diamond lattice with a unit cell consisting of three sites denoted as 'u' (up), 'c' (center) and 'd' (down) sites as sketched in Fig.~\ref{fig:flat_lattice}(a). 
The corresponding Hamiltonian for the k-th unit cell is given by
\begin{equation}
    \hat{H} = -J\sum_{k=1}^{L/3} ( e^{-i\phi} \hat{u}^{\dagger}_{k} \hat{c}_{k} + \hat{c}^{\dagger}_{k} \hat{d}_{k} + \hat{c}^{\dagger}_{k} \hat{u}_{k+1} + \hat{c}^{\dagger}_{k} \hat{d}_{k+1} + \text{H.c.}).
    \label{eq:Ham}
\end{equation}
Here, $\hat{u}^{\dagger}_{k},\; \hat{c}^{\dagger}_{k}$ and $\hat{d}^{\dagger}_{k}$ ($\hat{u}_{k},\; \hat{c}_{k}$ and $\hat{d}_{k}$) 
represent the fermionic creation (annihilation) operators for the k-th unit cell at 'u', 'c' and 'd' sites, respectively. The parameter $J$ represents nearest-neighbour coupling strength and $L$ represents the system size.
The hopping between the 'c' and 'u' sites within the unit cell is complex, acquiring a geometric phase $\phi$,  which corresponds to a total magnetic flux of $\phi/2\phi_0$ per plaquette in the lattice~\cite{PhysRevLett.111.185301,gaugefields2018}, where $\phi_{0} = h/e$ is the magnetic flux quantum.
We refer to the phase $\phi$ as the magnetic flux throughout this work. Since we are interested in transport properties, open boundary conditions are employed.
\begin{figure}[t!]
    \centering
    \includegraphics[width=1\linewidth]{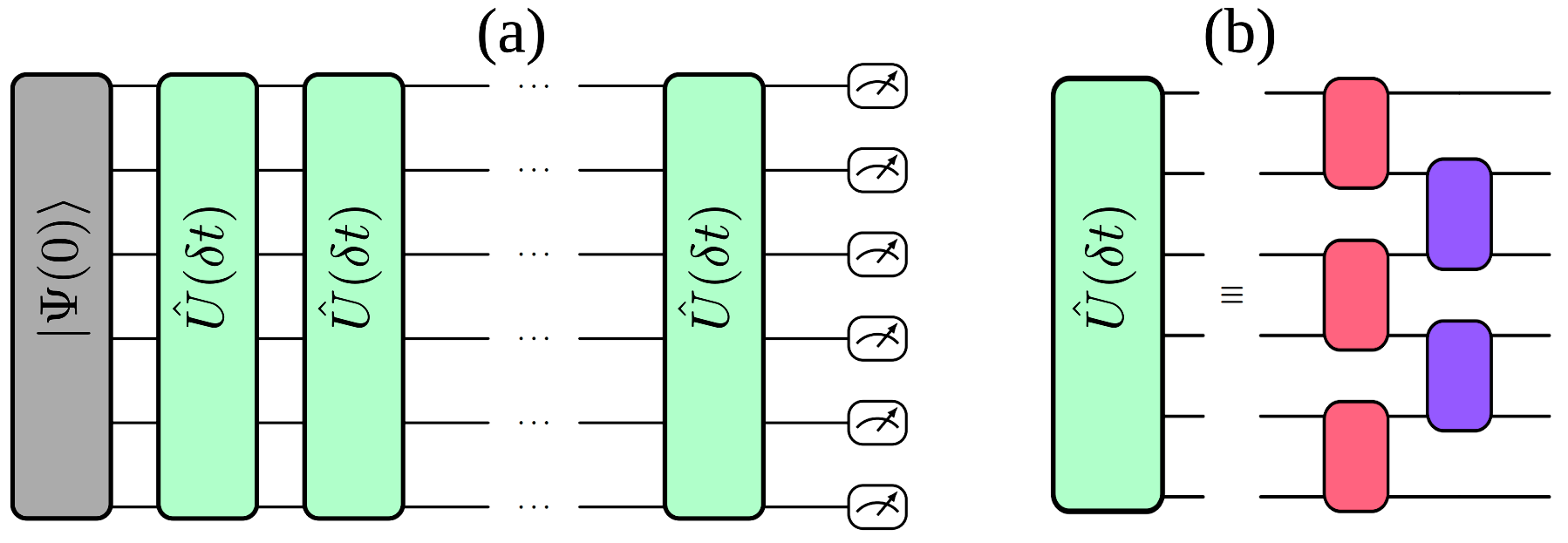}
    \caption{
    (a) The figure shows time evolution operator $\hat{U}(t)$ applied to the initial state $\ket{\psi(0)}$ where $\hat{U}(t)$ is discretized into discrete steps  (b) The unitary operators $\hat{U}(\delta t)$ are decomposed into two-qubit unitary operators.}
    \label{fig:ops}
\end{figure}
Furthermore, we note that FB and ABF lattices can be realized by tuning the flux $\phi$ to $0$ and $\pi$, respectively. At $\phi=0$, the band structure exhibits energy dispersion with respect to the quasi-momentum $q$, corresponding to a FB lattice with both flat and dispersive bands, as shown in Fig.~\ref{fig:flat_lattice}(b). However, when the flux is tuned to $\phi=\pi$, all bands become perfectly flat, resulting in an ABF lattice, as shown in Fig.~\ref{fig:flat_lattice}(c).

Our studies focus on the unitary time evolution of the time independent Hamiltonian $\hat{H}$, described as $|\psi(t)\rangle = e^{-i\hat{H}t} |\psi(0)\rangle$, where $|\psi(0)\rangle$ represents some initial state. 
To comprehend the dynamics, we compute the on-site particle density defined as, $n_i(t) = \langle \psi(t) | \hat{n}_i |\psi(t)\rangle$, where $\hat{n}_i$ denotes the density operator at lattice site $i$.
To simulate the Hamiltonian given in Eq.~\ref{eq:Ham} on a digital quantum computer, 
we first utilize the Jordan-Wigner (JW) transformation~\cite{qubitmap} to map it to map it onto a qubit representation. In this work we focus on the flux values $\phi=0$ (FB) and $\phi=\pi$(ABF). For these specific values, the complex hopping phases are real ($\eta = \pm 1$), and the corresponding Hamiltonian in terms of Pauli operators can be written as:
\begin{align}
    \hat{H} = -\frac{J}{2} \sum_{k=1}^{L/3} \Big[ \eta (X_k^u X_k^c + Y_k^u Y_k^c) + X_k^c X_k^d + Y_k^c Y_k^d  + X_k^c X_{k+1}^u + Y_k^c Y_{k+1}^u + X_k^c X_{k+1}^d + Y_k^c Y_{k+1}^d \Big],
\end{align}
where, $X_k^i$ and $Y_k^i$ are the Pauli operators acting on the respective qubits corresponding to the $i$'th ($i \in u, c, d$) site in the $k$-th unit cell. The parameter $\eta = +1$ for the FB ($\phi = 0$) case and $-1$ for the ABF ($\phi = \pi$) case.
The transformed Hamiltonian is now suitable for efficient simulation on current NISQ-era digital quantum computers. 
The implementation on the IBM Quantum quantum computer follows a structured, multi-stage approach.
First, we represent qubits as lattice sites, where $\ket{0}$ denotes an unoccupied site and $\ket{1}$ an occupied site. The initial state is prepared by applying NOT gates to the appropriate qubits, thereby encoding the desired particle distribution in Fock space via occupation number states.
We then discretize the total evolution time into $r$ uniform intervals of step size $\delta t = \frac{t}{r}$ and total evolution $\hat{U}(t)$ is realized by successively applying the single-step operator $\hat{U}(\delta t)$ on the initial state $\ket{\psi(0)}$ (see Fig.~\ref{fig:ops}(a)), such that
\begin{equation}
    |\psi(t)\rangle = \hat{U}(t) \ket{\psi(0)} = e^{-i\hat{H}t} \ket{\psi(0)} = \Big(e^{-i\hat{H}\delta t}\Big)^r \ket{\psi(0)} = \Big(\hat{U}(\delta t)\Big)^r \ket{\psi(0)}.
\end{equation}

To implement the single-step time evolution operator $\hat{U}(\delta t)$, we use the Suzuki-Trotter decomposition~\cite{Hatano2005}, which breaks down the evolution operator into a sequence of two-qubit unitary gates, as shown in Fig.~\ref{fig:ops}(b). In this study, we employ first-order Suzuki-Trotter decomposition. These unitary gates are further decomposed into hardware-native single- and two-qubit gates compatible with the IBMQ architecture~\cite{PhysRevA.69.032315}.
After constructing the quantum state $\ket{\psi(t)}$ at each time step, we perform measurements in the computational $z$-basis to extract the on-site particle density, defined as $n_i(t) = \langle \psi(t) | (\frac{1-\hat{\sigma}^z_i}{2}) |\psi(t)\rangle$ for each qubit. This allow us to study the evolution of particles across the lattice over time.

\section{Quantum simulation on IBM quantum computers}

In this study, we present simulation results using the 133-qubit IBMQ "Heron r1" processor from the "{\it{ibm\_torino}}" instance.
The error per logical gate (EPLG) is less than $1\%$, indicating high-fidelity two-qubit gate operations. Such a low EPLG mitigates cumulative errors in deep quantum circuits and enhances the performance of error-sensitive quantum algorithms~\cite{mckay2023}. 
To benchmark our findings, we compared results from IBMQ with results from exact diagonalization ("\text{Exact}") and an ideal quantum simulator ("{\text{Simulator}}") available through the Qiskit API~\cite{qiskit}. 
All quantum circuits were executed with $4096$ shots and a discrete time step of $\delta t = 0.1$, ensuring sufficient statistical accuracy for validating our results (see Appendix~\ref{shots_count}).
To mitigate hardware-induced errors, we employ the highest transpilation settings (optimization level = 3), which applies aggressive gate simplification and optimal qubit layout selection during circuit compilation to improve execution fidelity.  Additionally, we implement a post-selection protocol based on total particle number conservation~\cite{PhysRevA.98.062339, PhysRevLett.122.180501,Giri_2025}. Measurement outcomes violating the particle number constraint are discarded, and the remaining probabilities are renormalized. Details of the implementation on IBMQ given in Appendix~\ref{A:methods}. These procedures improve fidelity and ensure more reliable simulation of Hamiltonian dynamics on IBM Quantum hardware.
\begin{figure*}[h!]
    \centering
    \includegraphics[width=0.96\linewidth]{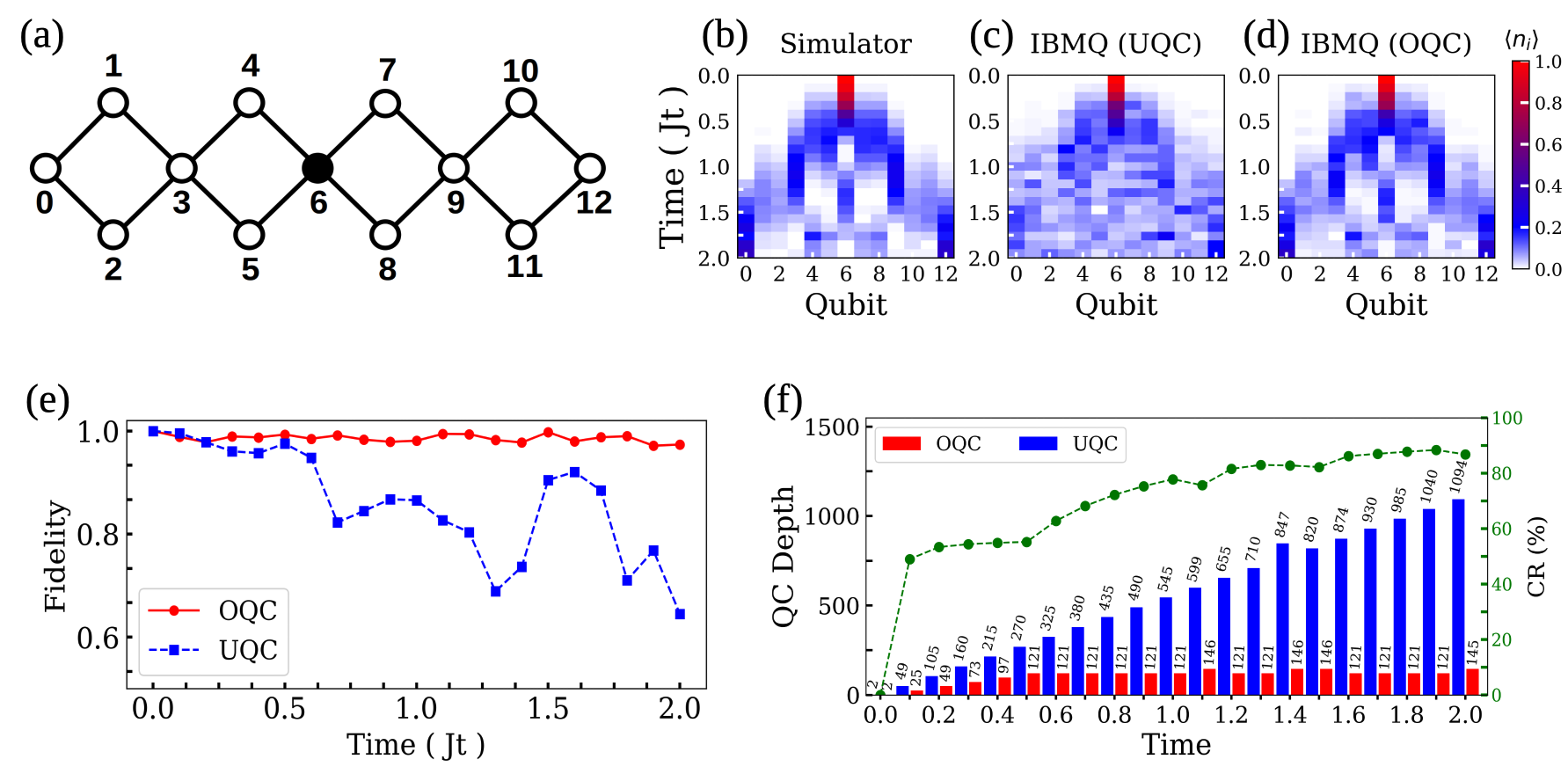}
    \caption{(a) Schematic of the 13-site flat-band lattice with the particle initially localized at the center and corresponding initial state is $\ket{\psi_0} = c_{6}^{\dagger}\ket{vac}$. (b) Density evolution of the particle obtained from the Simulator. (c) Density evolution of the original, unoptimized quantum circuits (UQCs) executed on IBMQ. (d) Density evolution of the optimized quantum circuits (OQCs) executed on IBMQ. The color bar indicates the particle density $\langle n_i \rangle$ at each site. (e) Fidelity over time for UQCs (blue squares) and OQCs (red circles) executed on IBMQ, benchmarked against the simulator. (f) Circuit depths of UQCs (blue bars) and OQCs (red bars) at different evolution times on IBMQ. The green dashed line indicates the compression ratio (CR) in \%, highlighting the significant reduction in circuit depth achieved through optimization.}
    \label{fig:compare_density}
\end{figure*}
\section{Compress the Quantum circuits using Tensor Network}

Simulating the long-time evolution of a Hamiltonian on NISQ-era digital quantum computers remains a significant challenge. Specifically, when time evolution is implemented using the Suzuki–Trotter decomposition, the circuit depth increases linearly with simulation time due to the sequential application of unitary operations. As the depth increases, the system becomes more susceptible to noise and decoherence, which degrade performance and lead to unreliable or unphysical results.
To overcome this, we propose to compress the quantum circuit on a classical computer and then run the compressed circuit on NISQ devices.
To illustrate the effects, we study the quantum walk of a particle initially localized at the center (6th site) of a 13-site flat-band lattice as sketched in Fig.~\ref{fig:compare_density}(a) and evolve under the Hamiltonian given in Eq.~\ref{eq:Ham} with $\phi=0$.
In Fig.~\ref{fig:compare_density}(b) we present the density evolution of the walker obtained from the Simulator, where a clear and coherent evolution pattern is observed. While in Fig.~\ref{fig:compare_density}(c) we show the results from IBMQ where unoptimized quantum circuits (UQCs) are used. It is clear that the result from the IBMQ with UQCs quickly loses fidelity compared to the ideal simulation.
It fails to maintain coherence and producing negligible meaningful results after only a few time steps. This highlights the need for advanced circuit optimization techniques to mitigate noise and decoherence, thereby enabling more accurate quantum simulations on NISQ devices.

The basic idea of circuit compression is to compress the circuit into a high fidelity effective circuit with a bounded depth. When the compressed circuit is sufficiently shallow for practical execution, it becomes possible to mitigate hardware noise and enable long-time evolution.
Several methods have been developed to optimize and reduce the depth of quantum circuits ~\cite{Nam2018,Bassman2022,Jones2022,quantum2025}. In this context, tensor network techniques have emerged as powerful tools for quantum circuit simulation, optimization, and simplification~\cite{markov2008simulating,Gray2021hyperoptimized,N.Cáliz2025,Berezutskii2025}.
In this work, we adopt an efficient approach based on the tensor network framework, which provides a compact, parametrized representation of quantum circuits. This enables precise optimization of the parameters of the compressed ansatz circuit with bounded depth. In particular, we have utilized the QUIMB library to optimize and compress the quantum circuit~\cite{gray2018quimb},
and the details of the optimization method is presented in Appendix~\ref{apndx-optimization}.
In Fig.~\ref{fig:compare_density}(d), we show the evolution of the optimized quantum circuits (OQCs) executed on IBMQ. We observe that the dynamics aligns closely with the ideal quantum simulator. 
To quantify the agreement between the Simulator and IBMQ results, we compute the fidelity at time $t$ as the square of the Bhattacharyya coefficient (BC)~\cite{BD,Hosseiny2024}:
\begin{equation}
    \centering
    F(t) =  BC^{2}(t),
    \label{eq:fidelity}
\end{equation}
where $BC(t) = \sum_{i} \sqrt{P_{\text{ibm},i}(t)\; P_{\text{sim},i}(t)}$. Here, $P_{\text{sim},i}(t)$ and $P_{\text{ibm},i}(t)$ denote the occupation probabilities of the particle at site $i$ and time $t$, as obtained from the Simulator and IBMQ hardware, respectively.
This basically quantifies how accurately IBMQ reproduces the ideal evolution. In Fig.~\ref{fig:compare_density}(e), we show a fidelity comparison between the OQCs and UQCs executed on IBMQ. We find that the OQCs consistently achieve a fidelity of approximately$\sim 97\%$ for all total evolution time. In contrast, the UQCs exhibit a rapid degradation in fidelity over time. Since the post-selection protocol and the transpilation settings (optimization level=3) are identical for both circuit types, the sustained high fidelity of the OQCs confirms that circuit compression is the dominant factor enabling accurate long-time simulation on the IBMQ hardware.
Moreover, in Fig.~\ref{fig:compare_density}(f), we present the quantum circuit depth ($D$) at each time step for both UQCs and OQCs. A substantial reduction in circuit depth is clearly achieved through optimization. We also calculate the compression ratio (CR), defined as $\text{CR}~(\%) = \left( \frac{D_{\text{UQC}} -D_{\text{OQC}}}{D_{\text{UQC}}} \right) \times 100,$ and observe that the circuit compression exceeds $80\%$ (see green dashed lines). These results highlight the significance of tensor network based variational optimization in achieving efficient and reliable execution of quantum circuits on IBMQ.

\section{Results}

In the following, all IBMQ results correspond to simulations of the OQCs on the IBMQ device, unless stated otherwise.
\begin{figure}[h!]
    \centering
    \includegraphics[width =1\columnwidth]{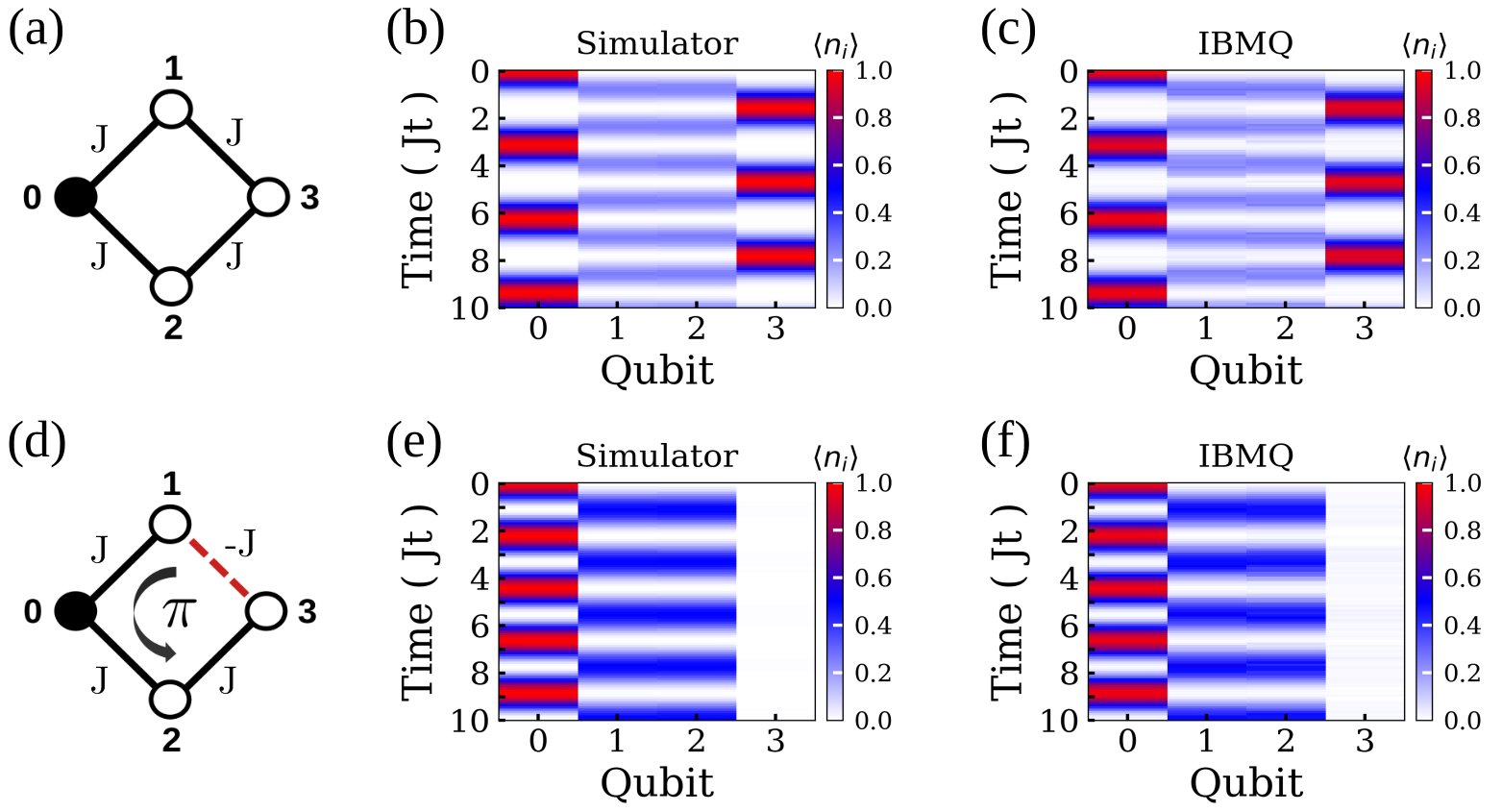}
    \caption{(a) Schematic of the 4-site FB lattice with uniform hopping ($J$), where the particle is initially localized at site '0'. Figures (b) and (c) show the density evolution on the Simulator and IBMQ, respectively. (d) Schematic of the ABF lattice by reversing the hopping ($-J$) on one of the links. Figures (e) and (f) show the density evolution on the Simulator and IBMQ, respectively. The colour bars indicate the particle density $\langle n_i \rangle$ at each site.}
    \label{fig:loc_dyn}
\end{figure}
\begin{figure*}[ht!]
    \centering
    \includegraphics[width =1.00\linewidth]{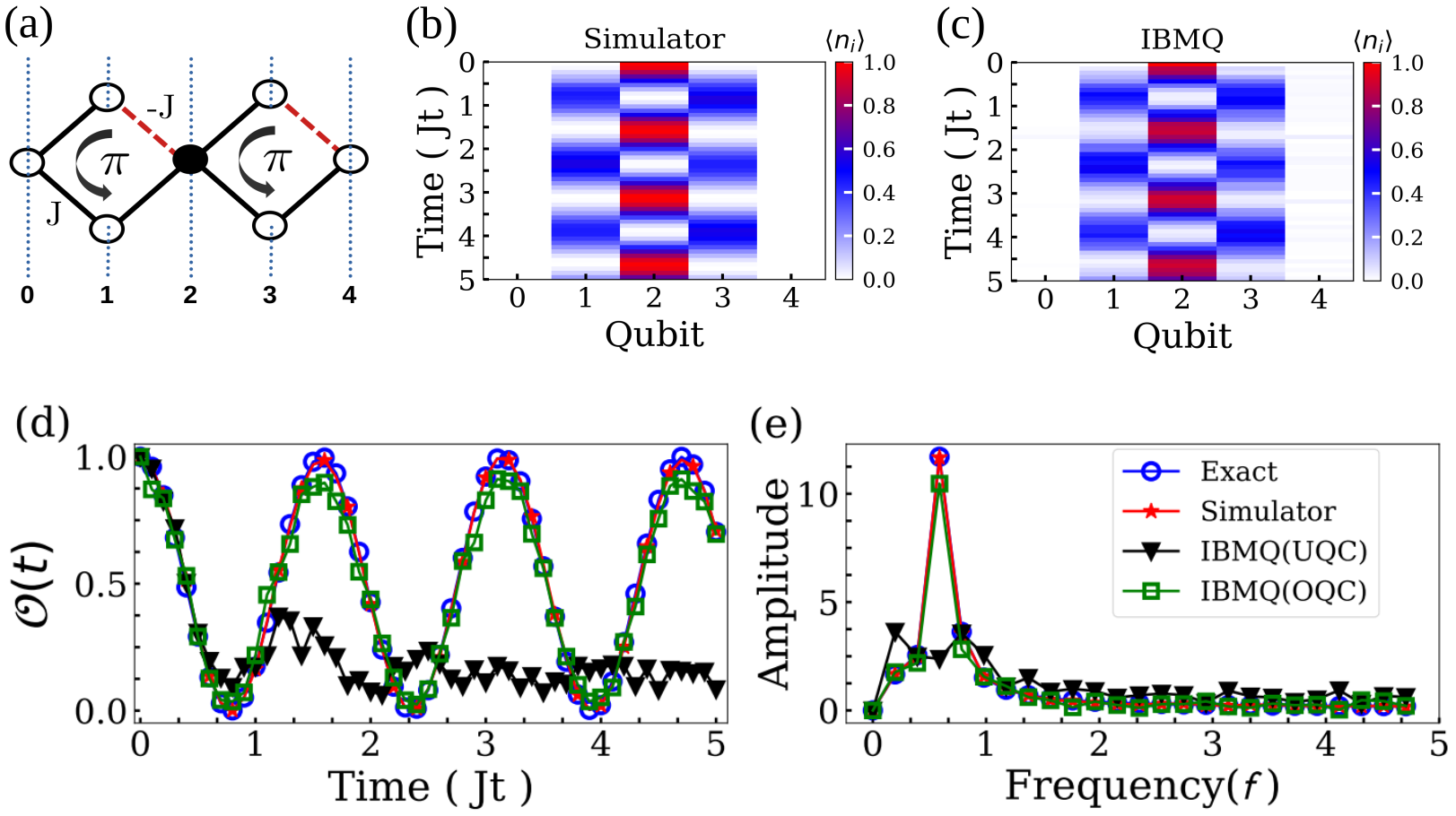}
    \caption{(a) Schematic of the ABF lattice consisting of two unit cells and the red dashed line indicates the reversal of hopping ($-J$) in the link. Figures (b) and (c) show the density evolution on the Simulator and IBMQ, respectively. The color bar represents the particle density $\langle n_i \rangle$ at each site.(d) The Figure shows the overlap $\mathcal{O}(t)$ define in Eq.~\ref{eq:overlap} as a function of time. (e) The Figure shows the Fast Fourier Transform (FFT) of the overlap, $\mathcal{O}(t)$.}
    \label{fig:trapping_dynamics}
\end{figure*}
\subsection{Single Particle Dynamics}
In this section, we investigate the single-particle quantum walk (QW) on a diamond lattice.
The FB and ABF lattices are realized by tuning the flux to $\phi=0$ and $\phi=\pi$ 
as sketched in Fig.~\ref{fig:loc_dyn}(a) and Fig.~\ref{fig:loc_dyn}(d) respectively. 
We first  focus on a single plaquette to effectively capture the localization behaviour of the entire lattice.
We perform the quantum walk with the particle initially localized at site "0". 
The results obtained from the simulator and IBMQ are presented respectively
in Fig.~\ref{fig:loc_dyn}(b,c) for the FB lattice and in  Fig.~\ref{fig:loc_dyn}(e,f) for the ABF lattice. 
We find that in the FB lattice, the particle becomes fully delocalized across all sites of the plaquette.
In contrast, in the ABF lattice, the walker remains confined to a subset of the plaquette. 
This distinction arises from the different band structures of the two lattices. 
Specifically, the FB lattice contains one flat band and two dispersive bands 
that touch at the two ends of the Brillouin zone as shown in Fig.~\ref{fig:flat_lattice}(b).
As a result, any initial state has some contributions from the dispersive bands. In the case of a single plaquette, the wavefunction undergoes constructive interference at sites '0' and '3' during the time evolution, resulting in coherent oscillations across the plaquette, as shown in Fig.~\ref{fig:loc_dyn}(b) and (c), obtained from the Simulator and IBMQ, respectively.
In contrast, when the lattice is extended, the wavefunction becomes fully delocalized throughout the entire lattice as shown in Fig.~\ref{fig:compare_density}. This delocalization results from complex interference and a continuous energy spectrum, which disrupts the regular oscillations observed in a single plaquette.
On the other hand, the ABF lattice consists entirely of flat bands as shown in Fig.~\ref{fig:flat_lattice}(c).
Consequently, any initial state is a superposition of CLSs. These CLSs arise due to destructive interference, which prevents the wavefunction from spreading beyond a limited region of the lattice. 
As a result, in the QW, the walker's movement is confined to sites up to '2' due to destructive interference of the wavefunction at site '3', as shown in Fig.~\ref{fig:loc_dyn}(d). This confinement is a clear demonstration of the AB caging effect~\cite{PhysRevLett.81.5888,PhysRevLett.83.5102,PhysRevLett.129.220403}. Notably, even in the extended lattice, the dynamics remain essentially unchanged. This is because all flat band structure and persistent destructive interference continue to restrict the particle's motion to a limited number of sites, preventing it from spreading across the extended lattice. Such a confinement effect has significant applications for particle trapping within the lattice.
In Fig.~\ref{fig:trapping_dynamics} we show the quantum walk of a particle initialized at the center of two plaquettes, where $\phi$ is set to $\pi$. 
We observe that  the walker's dynamics remain bounded during the evolution.
It exhibits near-perfect Bloch oscillation-type dynamics around the center, while the two edge sites remain inaccessible.
This behaviour highlights the potential for controlled quantum transport and suppression of qubit crosstalk.
To quantify the coherence properties of the qubits over extended periods, we calculate the overlap between the evolved state and the initial state.
The overlap at time $t$ is defined as 
\begin{equation}
    \mathcal{O}(t) = \sum_{i}n_{i}(0) n_{i}(t),
    \label{eq:overlap}
\end{equation}
where, $n_{i}(0)$ and $n_{i}(t)$ represent the density of the particle at site $i$ at time, $0$ and $t$, respectively. 
In Fig.~\ref{fig:trapping_dynamics}(d) we plot $\mathcal{O}(t)$ as a function of time. 
The key observation here is that the IBMQ results with OQCs exhibit periodic revivals, following closely the exact and simulator results.
Moreover, the revival amplitude of the overlap remains consistently high (>95\%) relative to the exact and simulator results.
This indicates that there is minimal decoherence in the system, attributable to the circuit compression.
In contrast,  IBMQ results with UQCs quickly deviated from the exact results. While one can still vaguely observe the periodicity, the amplitude revival is small.

We also calculate the Fast Fourier transform (FFT) of the overlap to quantify the oscillation frequency.
The FFT of the $\mathcal{O}(t)$ is defined as
\begin{equation}
    \text{FFT}(f) = \sum_{t=0}^{N-1} \mathcal{O}(t) e^{-2 \pi i f t / N},
\end{equation}
where $f$ is the frequency. Here $N$ corresponds to the total number of data points, which is equivalent to the total number of time steps. 
The FFT transforms time-domain signals into the frequency domain, allowing us to identify characteristic frequency that governs the system dynamics. 
This analysis also provides a framework to evaluate system noise, external perturbations, decoherence, and the overall fidelity of the quantum simulation. 
Due to periodic oscillation in the dynamics, this quantity defines the coherence nature of the system. 
If there is any decoherence in the system, the oscillation will decay with time and in the FFT one will observe a broader peak in stead of a sharp peak. 
In Fig.~\ref{fig:trapping_dynamics}(e) we present the FFT spectra for the exact calculation, the Simulator, 
and IBMQ hardware with UQCs and OQCs. We find that the FFT spectra for the exact calculation, the simulator, and the IBMQ with OQCs all exhibit a single sharp peak at $f \sim 0.59 J$, reflecting stable coherent oscillations. The IBMQ results with OQCs show a slightly reduced peak amplitude compared to the exact and simulator results, indicating minor coherence loss due to hardware imperfections.
In contrast, the IBMQ results with UQCs display a rapid loss of coherence within the first few time steps.
This result highlights both the effectiveness of our circuit optimization and the potential of the digital quantum computing platform to simulate complex quantum dynamics.

\begin{figure}[h!]
    \centering
    \includegraphics[width =1\linewidth]{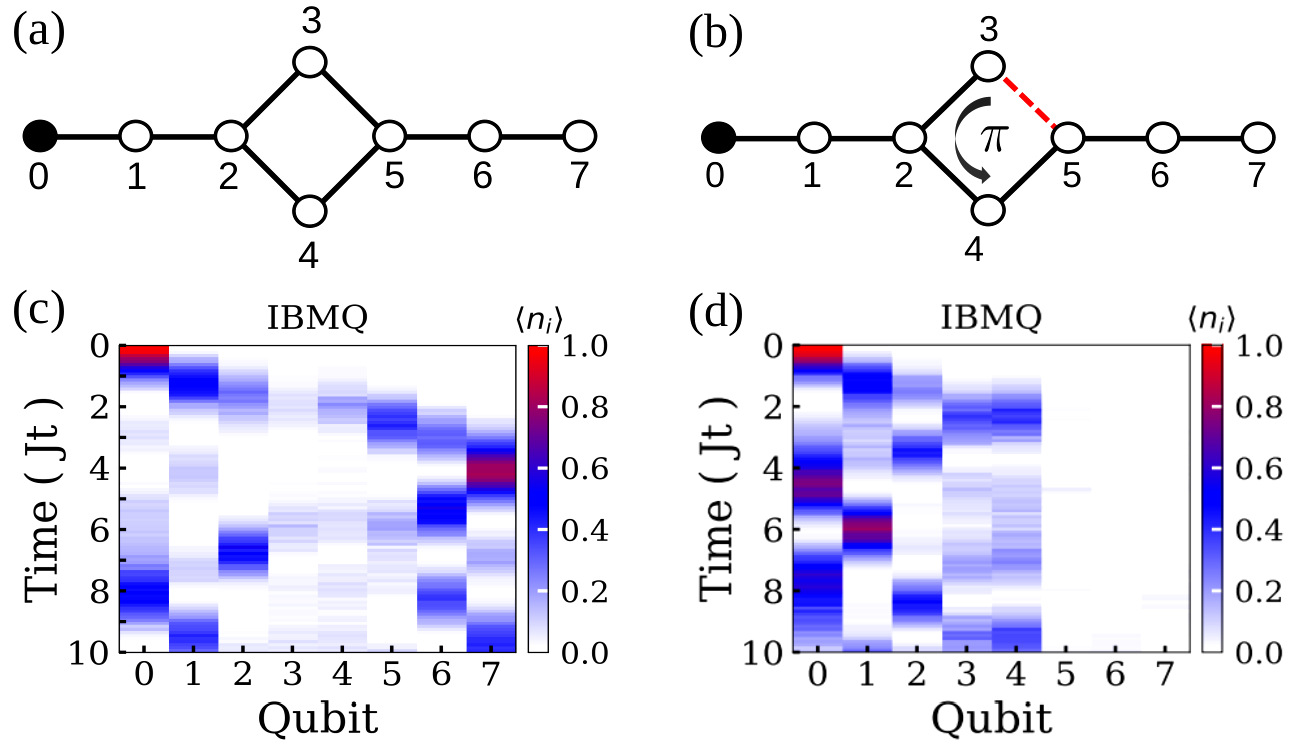}
    \caption{(a) Schematic representation of a flat-band lattice embedded in a one-dimensional lattice, with a particle initially localized at site 0. The corresponding density evolution is shown in (c). (b) The same lattice with an introduced $\pi$-flux, implemented by reversing the hopping ($-J$) on one edge of the plaquette. The corresponding density evolution is shown in (d).}
    \label{fig:flat_one}
\end{figure}
\subsection{Quantum Transport  }
After showing coherent quantum dynamics on the FB and ABF lattices, 
in this section, we show how these lattices can be utilized to control quantum transport. We embed an FB and ABF lattice within a one-dimensional chain as depicted in Fig.~\ref{fig:flat_one}(a) and \ref{fig:flat_one}(b), respectively.
We perform the quantum walk of a particle initially localized at the left edge (at site 0).
The goal is to examine its ability to traverse the embedded lattice and reach the right edge. 
In Fig.~\ref{fig:flat_one}(c), we present the IBMQ results for the FB lattice case. The walker undergoes a unidirectional quantum walk, propagating efficiently across the system and reaching the opposite edge.
In contrast, the IBMQ results for the case of ABF lattice are presented in Fig.~\ref{fig:flat_one}(d), where different behaviour is observed.
Initially the walker follows a unidirectional QW. But upon reaching site '2', the wavefunction splits symmetrically, leading to destructive interference at site '5'. 
Consequently, the walker fails to reach the opposite edge of the lattice. 
This behaviour confirms that the ABF lattice acts as a quantum transport switch since it effectively limit the propagation of the wavefunction. 
By reversing the hopping amplitude on one leg of the FB lattice one can switch between FB and ABF configurations. This provides a flexible approach to controlling quantum transport with applications in quantum computing and communication. 
As a quantum switch, it can direct the information flow in quantum networks and regulate qubit interactions. In quantum communication, it facilitates coherent state transfer, enhances quantum memory, and improves qubit isolation by minimizing crosstalk and decoherence. This approach paves the way for robust and scalable quantum computing by enabling optimized qubit connectivity.

\begin{figure}[t]
    \centering
    \includegraphics[width =1\columnwidth]{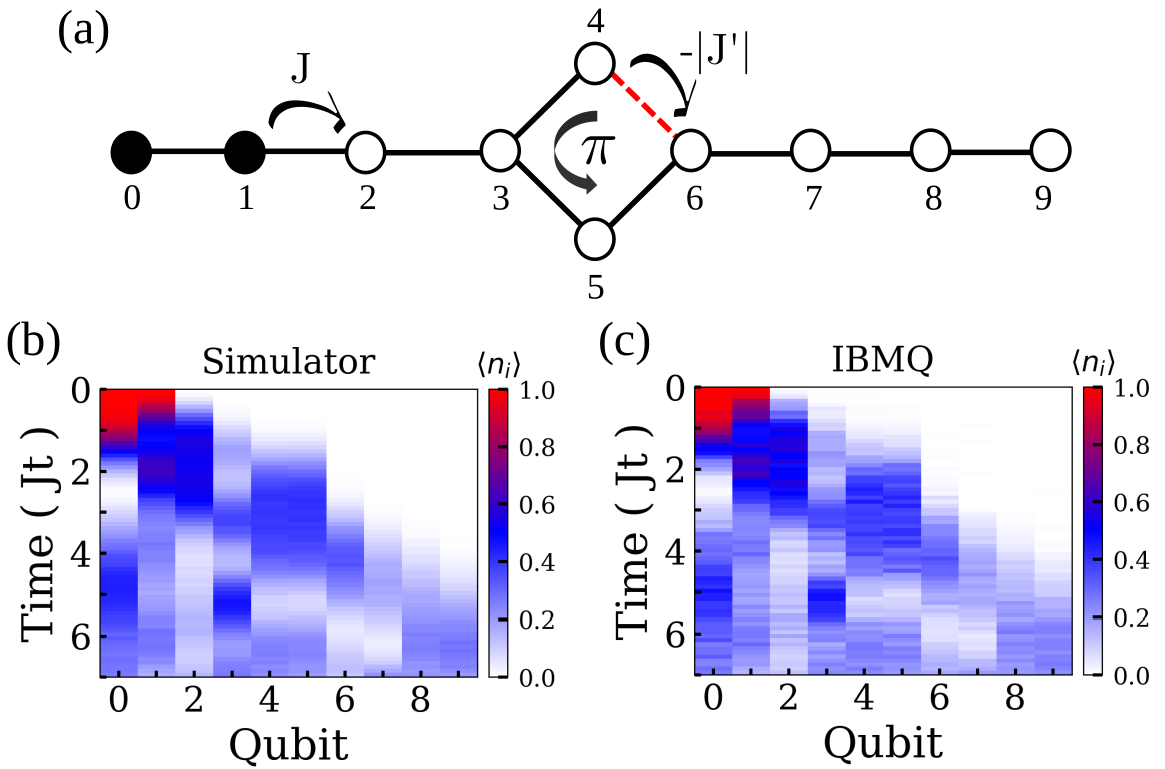}
    \caption{(a) Schematic representation of an all-bands-flat lattice embedded within a one-dimensional chain. The red dashed line indicates $\rm -|J'|$. Initially, two particles are localized at sites 0 and 1, respectively. Figures (b) and (c) show the time evolution obtained from an Simulator and IBMQ for the case where $\rm |J'| = |J| = 1$, respectively.}
    \label{fig:flat_two}
\end{figure}
\begin{figure*}[h!]
    \centering
    \includegraphics[width =1.00\linewidth]{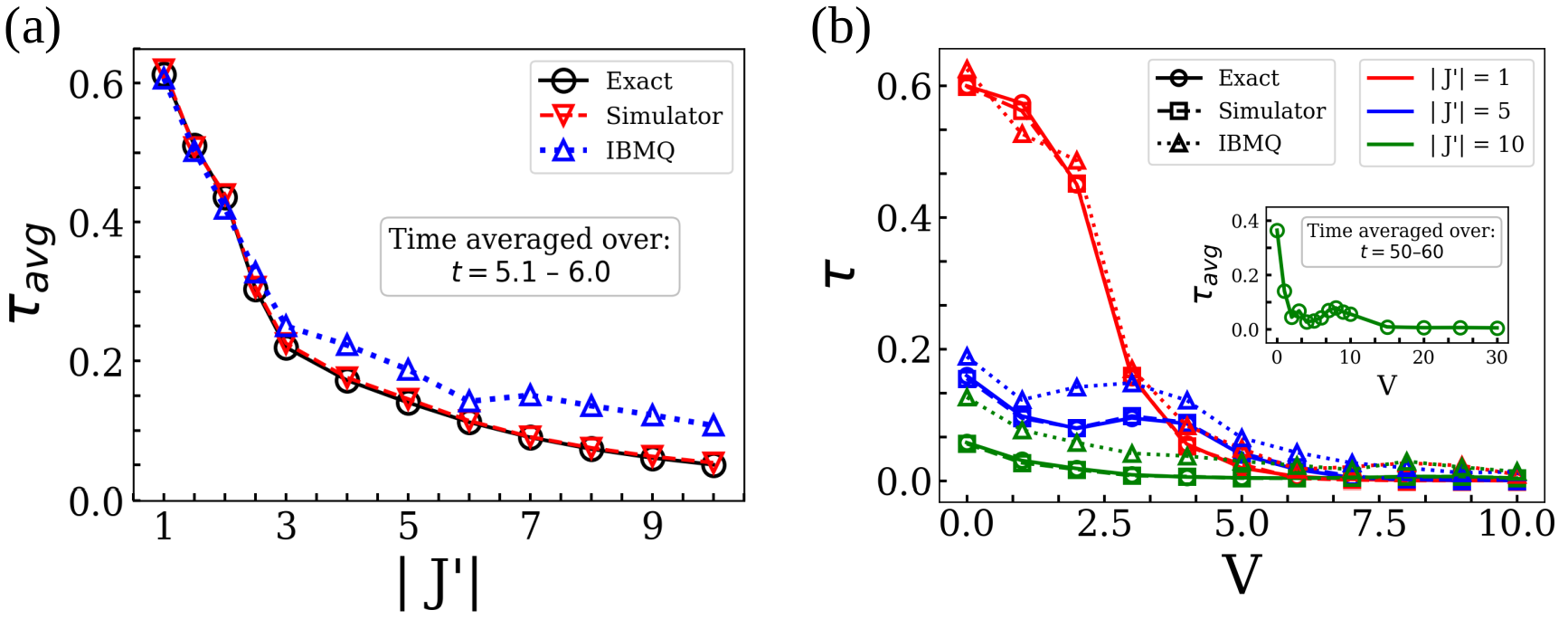}
    \caption{(a) The figure shows the time-averaged transmission coefficient $\tau_{avg}$, as defined in Eq.~\ref{eq:transmission}, as a function of $\rm |J'|$, obtained from exact calculations, Simulator, and IBMQ. (b) The figure presents the instantaneous transmission coefficient $\tau(t)$ at time $t = 6.0$ as a function of interaction strength for different values of $\rm |J'|$, based on data from exact calculations, an ideal quantum simulator, and IBMQ. The inset shows the long-time-averaged transmission $\tau_{avg}$, evaluated over the interval $50 \leq t \leq 60$ using exact calculations.
 } \label{fig:flat_two_tau}
\end{figure*}

Next, we investigate the behaviour of two non-interacting particles. 
We consider the configuration in which two particles are localized at sites '0' and '1' of the left edge in the ABF-embedded system as sketched in Fig.~\ref{fig:flat_two}(a). Results from simulator and IBMQ are presented in Fig.~\ref{fig:flat_two}(b) and (c) respectively, where $\rm |J'| = |J| = 1$. In contrast to the single-particle scenario, we observe that the wavefunction is able to traverse the ABF lattice.
Consequently quantum transport to the opposite end becomes allowed. This behaviour highlights how  collective dynamics can overcome the localization constraints typically imposed by the ABF structure, allowing for enhanced transport. 
Moreover, one can control transport by tuning the hopping strength $\rm |J'|$.
To capture this, we first define the instantaneous transmission at time $t$ as $\tau(t) \equiv \sum_{i=6}^{9} n_i(t)$, where $n_i(t)$ is the on site particle density at site $i$.
We then calculate the time-averaged transmission coefficient $\tau_{avg}$, which is defined as
\begin{equation}
    \tau_{avg} = \frac{1}{N}\sum_{t_i}^{t_f} \tau(t), 
    \label{eq:transmission}
\end{equation}
where $N = \frac{t_f - t_i}{dt}$, with $dt =0.1$. 
In Fig.~\ref{fig:flat_two_tau}(a), we show $\tau_{avg}$ as a function of $\rm |J'|$ over the time interval from $t_i = 5.1$ to $t_f = 6.0$. 
The results show that increasing $\rm J'$ suppresses transmission, providing a tunable mechanism for controlling transport within the system. With the help of circuit-compression, the results from IBMQ qualitatively follows the exact and simulator results. However, it exhibits slightly higher transmission values at larger $\rm J'$. This deviation emanates from the sensitivity of the quantum simulation to hardware noise as $\rm J'$ increases. As $\rm J'$ increases, it amplifies gate control errors, enhances susceptibility to decoherence due to faster wavefunction spreading across connected qubits, and increases trotterization errors during the evolution. These combined effects contribute to observable distortions in the quantum state, resulting in deviations in the transmission from the ideal behaviour even after optimization in the high~$\rm J'$ regime.

Finally, we investigate the effect of the interaction. We incorporate nearest-neighbour interactions by adding the interaction term $\hat{H}_{\text{int}}$ into the Hamiltonian, defined as
\begin{equation}
    \hat{H}_{\text{int}} = \frac{V}{4} \sum_k Z^{c}_{k} \left( Z^{u}_{k} + Z^{d}_{k} + Z^{u}_{k+1} + Z^{d}_{k+1} \right).
\end{equation}
Here $V$ denotes the interaction strength and $Z^{i}_{k}$ is the Pauli-$Z$ operator acting on site $i \in \{u, c, d\}$ within the $k$'th unit cell. 
The goal is to investigate whether modulation of the hopping amplitude $\rm J'$ can still effectively control quantum transport. 
As Fig.~\ref{fig:flat_two}(b) and (c) show, the wavefunction reaches the system’s boundary well before time $t=6.0$. Therefore, we compute the instantaneous transmission coefficient $\rm \tau(t)$ at time $ t = 6.0$ for different interaction strengths $ V $.
The results are presented in Fig.~\ref{fig:flat_two_tau} (b), which shows how transmission depends on both the interaction strength $V$ and the hopping amplitude $|J'|$.
First focus on the exact and simulator results. In all three cases, $\rm|J'| = 1$ (red), $5$ (blue), and $10$ (green), transmission decreases as the interaction strength $V$ increases. 
In the strong interaction regime ($V > 5$), two particles undergo a correlated quantum walk~\cite{Fukuhara2013,PhysRevA.90.062301}, which significantly slows the dynamics. As a result, transmission ceases to zero for all $\rm |J'|$. 
For $\rm|J'| = 1$, transmission, $\tau$ is rapidly suppressed with increasing interaction strength $V$ due to dephasing and scattering induced by the interaction. 
On the other hand, larger hopping amplitudes ($\rm|J'|$ = $5$ and $10$) lead to strong localization via AB caging, resulting in suppressed transmission even at $V=0$. Consequently, $\tau$ is further reduced when interactions are introduced.
We observe that results from IBMQ follow similar trends. Notably, the transmission at larger hopping strengths $\rm|J^\prime|$ is consistently higher in the IBMQ results, when compared to the simulator data. This is consistent with the behaviour observed in the non-interacting case. Moreover, for $\rm |J^\prime| = 5$ and $10$, transmission remains relatively high in the weakly interacting regime ($0 \leq V \leq 5$).
This is because the wavefunction will extend over a larger number of qubits and this increased delocalization enhances the phase decoherence on real hardware. In contrast, in the strongly interacting regime ($V \geq 5$), the wavefunction becomes confined to fewer sites as the particles undergo correlated quantum walks. 
As a result, transmission is significantly suppressed, and the deviation in $\tau$ between the IBMQ and simulator data becomes small. 

For more concreteness, we examine the long-time dynamics by computing the time-averaged transmission coefficient, $\tau_{\text{avg}}$, over the interval $50 \leq t \leq 60$ at $|J'| = 10$, varying interaction strengths. The results from exact calculations are shown as a green line with open circles in the inset of Fig.~\ref{fig:flat_two_tau}(b). The transmission exhibits a non-monotonic dependence on $V$. Initially, $\tau_{\text{avg}}$ decreases rapidly as $V$ increases, indicating strong suppression of transport. But, a minor revival in $\tau_{\text{avg}}$ is observed in the range $6 < V \leq 10$, likely due to resonant tunneling as the interaction strength approaches the hopping amplitude $|J'| = 10$. In this regime, interaction-assisted processes can momentarily enhance particle transport before strong interactions suppress it completely.

In summary, these findings confirm that even in the presence of interactions, the hopping amplitude $|J'|$ remains an effective and tunable parameter for controlling particle transport in flat-band engineered systems. This has several important implications for quantum technologies. It can be utilized to create quantum switches and filters that manage the propagation of quantum states based on adjustable system parameters, which is crucial for routing quantum information while minimizing crosstalk. Furthermore, the strong localization properties of ABF configurations make them ideal for qubit isolation, helping to reduce error rates and mitigate decoherence. In quantum memory, the ability to trap quantum information within localized states significantly enhances coherence times.

\section{Discussion}

We study QWs on FB and ABF lattices using a digital quantum computer. FB and ABF lattices can be realized within the diamond chain with precise tuning of the hopping amplitude. The ABF lattice is realized by introducing $\pi$ flux per plaquette. It is known that simulating long-time evolution on NISQ digital quantum platforms presents challenges due to the linear increase in circuit depth. Hardware noise and decoherence can easily result in unphysical outcomes. To overcome these limitations, we implement tensor network based variational optimization, which significantly compresses circuit depth and consequently reduces the effects of noise and decoherence.  This approach enables the efficient execution of long-time quantum evolution while ensuring the practicality of circuit implementation on current NISQ quantum computers. 
We begin by studying QWs on a single-plaquette FB lattice, where we observe coherent oscillation. In contrast, extended FB lattices exhibit complete wavefunction delocalization and ballistic transport - stemming from complex interference patterns and a continuous energy spectrum. The ABF lattice, on the other hand, exhibits strong localization due to destructive interference, demonstrating Aharonov-Bohm (AB) caging. Even in the extended ABF lattice, the walker dynamics remain confined within a few sites, showing Bloch oscillation-type dynamics. We also show that the optimized quantum circuit maintains a consistently high wavefunction overlap. A well-defined peak is observed in the FFT analysis, validating the effectiveness of tensor network optimization. We note that the transition between FB and ABF configurations can be achieved by simply reversing the sign of the hopping amplitude on a single plaquette edge, offering a straightforward mechanism for quantum control in information processing applications. 
Furthermore, we show how the FB and ABF lattice can be used to control the quantum transport by integrating them within a one-dimensional structure. The ABF lattice acts as a quantum switch that regulates particle movement. Interestingly, in the case of two interacting particles, the ABF plaquette alone is insufficient to restrict transport; additional tuning of the hopping amplitude is required to control the dynamics.

In summary, our results show that quantum simulations of flat-band physics are possible on current near-term quantum devices. However, several challenges remain - including the scaling to larger systems with interactions and disorder~\cite{PhysRevA.108.L010201,PhysRevB.82.104209,PhysRevB.107.245110}. Ongoing advances in quantum hardware and circuit optimization methods, such as adaptive variational quantum eigensolvers (VQE)~\cite{Grimsley2019}, the Quantum Approximate Optimization Algorithm (QAOA)~\cite{PhysRevResearch.7.023165}, and advanced unitary synthesis~\cite{PhysRevLett.118.010501,quantum2025}- are rapidly improving the performance of quantum simulations. These improvements are expected to simulate complex physical systems - such as quantum materials~\cite{PhysRevX.12.040501,Head-Marsden2021}, many-body dynamics~\cite{RevModPhys.80.517,PhysRevB.107.L220201}, spin liquids~\cite{Savary_2017}, state transfer~\cite{state_transfer} and quantum phase transition~\cite{PhysRevA.78.042105,PhysRevResearch.3.033199,PhysRevB.100.184417}. Such systems lie at the critical interface of quantum information science and condensed matter physics.
\section*{Acknowledgements}
We acknowledge the support by National Science and Technology Council (NSTC) of Taiwan through Grant No. 114-2119-M-007-013. MKG acknowledges support for this research, which was partially funded by the Ministry of Education, Singapore through Tier 2 Grant MOE-T2EP50123-0021. Pochung Chen acknowledges the support by Taiwan Centers of Excellence (TCE), Ministry of Education, Twiwan.
\begin{appendices}
\appendixnumbering
\section{Implementation on IBMQ}
\label{A:methods}
\subsection{Initial State Preparation}
In this study, we simulate the time evolution of an initial quantum state using the IBM quantum computer. The default initial state is the vacuum state, with all qubits initialized in $\ket{0}$. To simulate a particle localized at a specific site, we prepared the initial state by applying NOT gates to the appropriate qubit(s). For example, to prepare the state $\ket{10100}$ on a five-qubit system, we apply Pauli-$X$ gates to the first and third qubits.

\subsection{Time evolution}
Once the initial state is prepared, we then need to implement the time evolution operator $\hat{U}(t)$. For this, we discretize the total evolution time $t$ into $r$ small time steps of size $\delta t = t/r$. The full evolution operator is then expressed as a product of evolution operators for each time step:
$\hat{U}(t) = \left[\hat{U}(\delta t)\right]^r = \underbrace{\hat{U}(\delta t) \cdot \hat{U}(\delta t) \cdots \hat{U}(\delta t)}_{r\ \text{times}},$ 
as shown in Fig.~\ref{fig:ops}(a). By repeatedly applying $\hat{U}(\delta t)$, we evolve the system to the desired final time $t$. 
To implement each $\hat{U}(\delta t)$ on the IBMQ device, we approximate the time evolution operator using the Suzuki–Trotter decomposition~\cite{hatano2005finding}. For a Hamiltonian of the form 
\begin{equation}
H = \sum_{i,\in,\text{odd}} h_{i,i+1} + \sum_{i,\in,\text{even}} h_{i,i+1} = H_o + H_e,
\end{equation}
the evolution operator can be expressed as
\begin{equation}
\hat{U}(\delta t) = e^{-iH\delta t} = e^{-iH_o \delta t} e^{-iH_e \delta t} + \mathcal{O}(\delta t^2).
\end{equation}
This corresponds to the first-order Suzuki–Trotter decomposition, which breaks the Hamiltonian into non-commuting even and odd components. This structure allows us to decompose the evolution operator into a sequence of two-qubit gates acting on nearest-neighbour qubits, as shown in Fig.~\ref{fig:ops}(b).

It is important to note that although higher-order decompositions more effectively reduce Trotter errors, they typically require additional layers of  two-qubit gates, which increases the overall circuit depth and thereby makes the circuit more susceptible to noise and decoherence on current quantum hardware. To balance the trade-off between Trotter error and hardware-induced noise, we adopt the first-order decomposition with a time step size of $\delta t = 0.1$.

At final step, we have to further decompose the two-qubit operators into the native quantum gates, which are RX, RY, RZ, CNOT and others - supported by the hardware~\cite{PhysRevA.69.032315,Smith2019}.

Under ideal conditions, this procedure yields the intended time-evolved state. However, current NISQ devices are error-prone and the quality of the evolved state degrades significantly after just a few time steps. To address this limitation, we employ a tensor network based circuit optimization, which will be discussed in the next section.
\begin{figure*}[th!]
    \centering
    \includegraphics[width=1.0\linewidth]{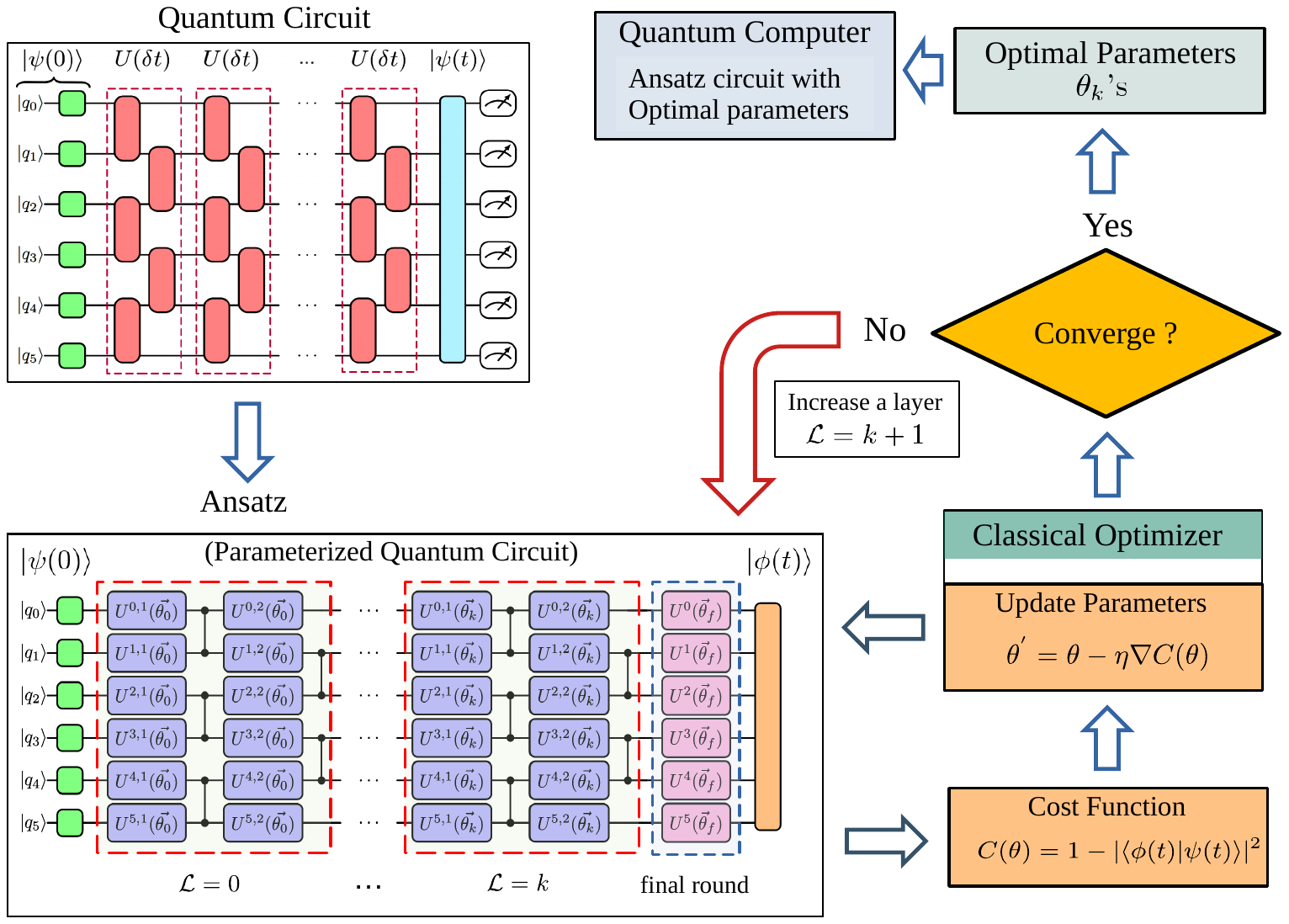}
    \caption{ A schematic diagram of the optimization process of the quantum circuit.}
    \label{fig:optimization}
\end{figure*}
\subsection{Quantum circuit optimization}
\label{apndx-optimization}
In the optimization process, our goal is to closely approximate the time-evolved quantum state $\ket{\psi(t)}$.  To do this, We construct a quantum circuit corresponding to the target time evolution operator and apply it to the initial state $\ket{\psi(0)}$. The evolved state is then represented as a tensor network, allowing for efficient parametrization and manipulation. To approximate this state, we construct a variational ansatz circuit using single and two-qubit nearest-neighbour random unitaries, denoted as $U(\theta_i)$, to form a parametrized quantum circuit within the tensor network framework that consists of $k$ layers of $U(\theta_i)$, as illustrated in Fig.~\ref{fig:optimization}. This parametrized circuit performs as a flexible model for simulating evolution of $\ket{\psi(t)}$. 
The ansatz is then applied to the initial state, generating the approximated evolved state $\ket{\phi(t)}$. To quantify the accuracy of this approximation, we calculate the fidelity-based loss function defined as 
\begin{equation}
    C(\theta) = 1 - |\langle \phi(t) | \psi(t) \rangle|^2,
\end{equation}
which quantifies the deviation between the exact and approximated states. 
Minimizing $C(\theta)$ confirms the high-fidelity approximated quantum state. The quantum circuit is optimized using the L-BFGS-B (Limited-memory Broyden-Fletcher-Goldfarb-Shanno with Box constraints) algorithm~\cite{optimization1,optimization2}, a quasi-Newton method that efficiently tunes the variational parameters while preserving unitary constraints. We use automatic differentiation to quickly and accurately calculate gradients, which helps speed up the optimization process. If the optimization saturates before reaching the desired fidelity, we increase the circuit depth $\mathcal{L}$ by one additional layer and repeat the optimization until convergence is achieved. This adaptive approach mediates the trade-off between circuit expressibility and computational efficiency, ensuring high-fidelity quantum state evolution with minimal computational cost. In this study we have use QUIMB library to compress the quantum circuits~\cite{gray2018quimb}.
This tensor network based variational optimization strategy allows for efficient compression of quantum circuits, significantly reducing the gate depth required for accurate simulation of quantum dynamics on near-term quantum hardware, as shown in Fig.~\ref{fig:compare_density}. 

We emphasize that the different system sizes considered in our study for the FB and ABF cases does not arise from limitations or fidelity degradation of the compression scheme. Rather, it reflects the distinct dynamical behaviour of the two Hamiltonians. In FB dynamics the wavefunction propagates across the lattice, which requires deeper circuits to accurately represent the evolving state. In contrast, ABF dynamics remain strongly localized, allowing accurate representation with shallower compressed circuits. Therefore, the chosen system sizes are sufficient to demonstrate the relevant physics in both cases.

\subsection{Measurement}
In our study, all gate operations and measurements were performed in the computational $z$-basis. Each quantum circuit corresponds to a time step was executed with $4096$ measurement shots, ensuring sufficient statistical accuracy for validating our results.

\subsection{Post-selection}
We study quantum dynamics in the occupation number basis, where the total particle number is preserved throughout the evolution. Since IBMQ provides outcome statistics for all sampled basis states, we apply a post-selection protocol to enforce particle-number conservation. Specifically, measurement outcomes that violate the total particle number constraint are discarded, and the probabilities of the remaining outcomes are renormalized. This post-selection procedure reduces the impact of hardware-induced errors and ensures that the analysis remains restricted to the physically relevant, number-conserving subspace. 
\begin{figure}
    \centering
    \includegraphics[width=0.9\columnwidth]{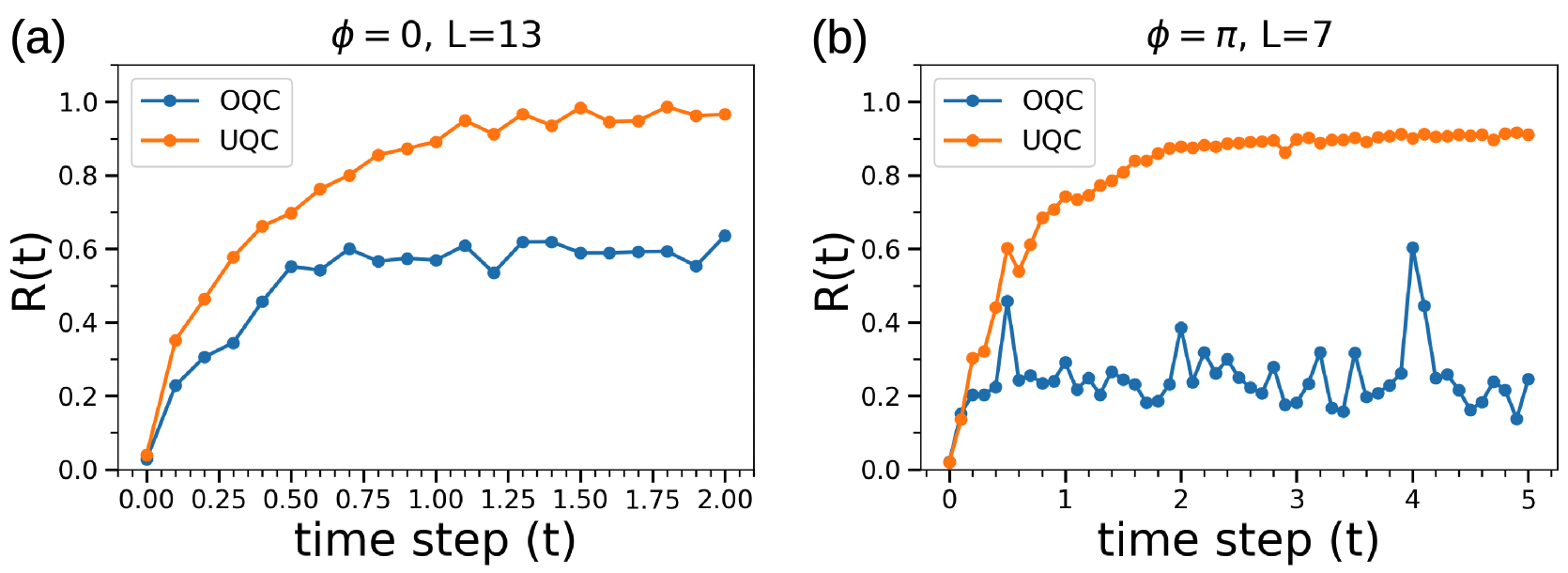}
    \caption{Rejection rate $R(t)$ as a function of time for unoptimized (UQC) and optimized (OQC) circuits. Figure (a) corresponds to system size $L = 13$, associated with the density evolution shown in Fig.~\ref{fig:compare_density}. Figure (b) corresponds to $L = 7$, associated with the density evolution shown in Fig.~\ref{fig:trapping_dynamics}.}
    \label{fig:rejection_plot}
\end{figure}

To quantify the cost of post-selection, we calculate the rejection rate $R(t)$ for each time step. The rejection rate is defined as 
\begin{equation}
R(t) = 1 - \frac{S_{ps}(t)}{S_{tot}(t)},
\end{equation}
where $S_{ps}(t)$ denotes the number of post-selected shots that satisfy the total particle number constraint and $S_{tot}(t)$ is the total number of measurement shots. We analyze $R(t)$ for both unoptimized (UQC) and optimized (OQC) circuits. The analysis is performed for system sizes $L = 13$ and $L = 7$, corresponding to the density evolution results from FB Hamiltonian shown in Fig.\ref{fig:compare_density} and from ABF Hamiltonian shown in Fig.~\ref{fig:trapping_dynamics}, respectively. The resulting rejection rates are presented in Fig.~\ref{fig:rejection_plot}(a) and (b), respectively. In all cases, the OQCs shows consistently lower rejection rates than the UQCs. Moreover, the $R(t)$ is noticeably lower in the ABF lattice (L=7) compared to the FB lattice (L=13), which can be attributed to the strict localization in the ABF system. These results indicate that post-selection remains practical and scalable for our study over the entire evolution window considered.
\end{appendices}
\begin{figure}[h!]
    \centering
    \includegraphics[width=1\columnwidth]{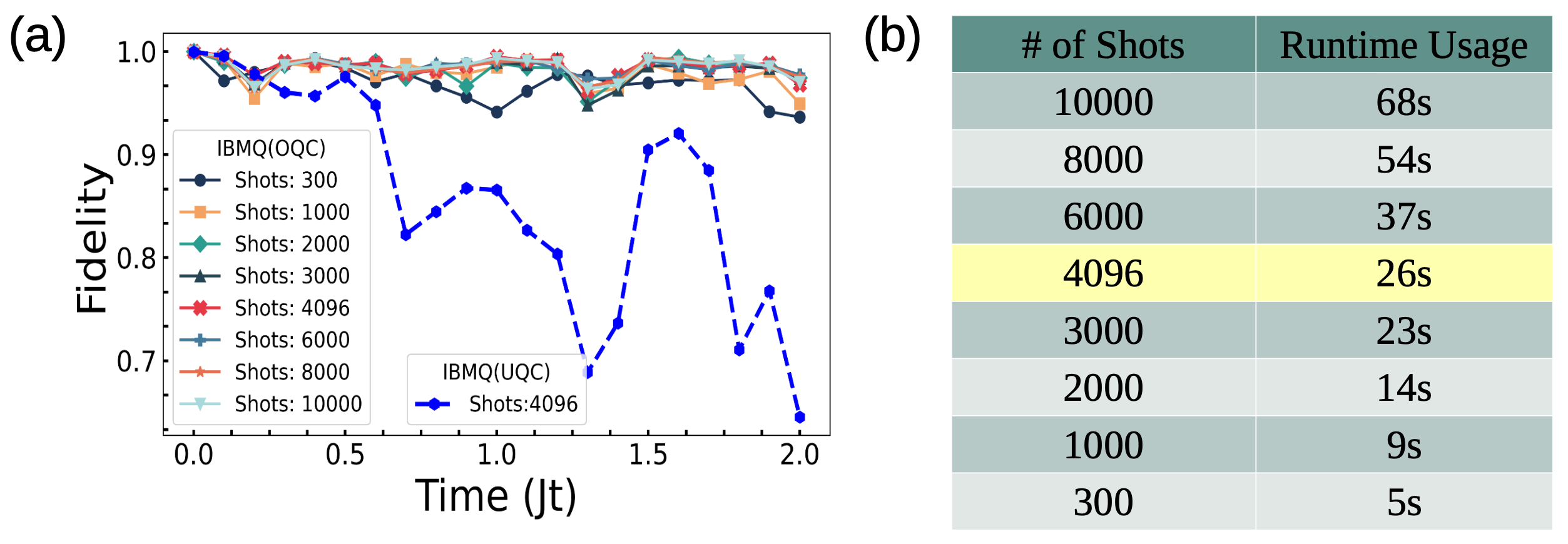}
    \caption{(a) Fidelity of unoptimized and optimized quantum circuits for various shot counts, executed on IBMQ. (b) Table showing the runtime usage corresponding to different shot counts.}
    \label{fig:shots_comparison}
\end{figure}
\begin{appendices}
\section{Fidelity and Runtime Analysis for Different Shot Counts on IBMQ}
\label{shots_count}
We present the fidelity, defined in Eq.~\ref{eq:fidelity}, for different numbers of measurement shots in Fig.~\ref{fig:shots_comparison}(a), with the corresponding density evolution shown in Fig.~\ref{fig:compare_density}. In Fig.~\ref{fig:shots_comparison}, the solid lines indicate results from optimized quantum circuits on IBMQ, whereas the dashed line corresponds to an unoptimized circuit on the same backend. Figure~\ref{fig:shots_comparison}(b) displays the runtime usage for different numbers of shots, showing that runtime increases with the number of shots and highlighting the importance of selecting a cost-effective balance between accuracy and computational resources. In our study, we chose 4096 shots, as the fidelity at this level is comparable to that at 10,000 shots while significantly reducing runtime and cost. This choice enables us to retain high fidelity while avoiding unnecessary computational overhead in larger-scale simulations.
\end{appendices}

\bibliography{references}

\end{document}